\documentclass[useAMS,usenatbib,twocolumn,floatfix]{mn2e}
\usepackage{natbib}
\usepackage{graphicx}
\usepackage{bm}
\usepackage{amsmath}
\usepackage{color}
\usepackage[varg]{txfonts}

\newcommand{\prd}{Phys.~Rev.~D}

\newcommand{\aj}{AJ}

\newcommand{\apj}{ApJ}
\newcommand{\apjl}{ApJL}
\newcommand{\apjs}{ApJS}
\newcommand{\jcap}{JCAP}

\newcommand{\mnras}{MNRAS}

\newcommand{\nat}{Nature}

\def\rref#1{equation (\ref{#1})}

\begin{document}

\title[Effect of DE perturbation on void]%
{Effect of Dark Energy Perturbation on Cosmic Voids Formation}

\author[Endo, Nishizawa \& Ichiki]
{Takao Endo$^{1}$\thanks{Email: endou.takao@a.mbox.nagoya-u.ac.jp},
Atsushi J. Nishizawa$^{1,2}$
and Kiyotomo Ichiki$^{1,3}$\\
$^{1}$ Graduate School of Science, Nagoya University, Aichi 464-8602, Japan,\\
$^{2}$ Institute for Advanced Research, Nagoya University, Aichi 464-8602, Japan,\\
$^{3}$ Kobayashi Maskawa Institute, Nagoya University, Aichi 464-8602, Japan}
\maketitle

\begin{abstract}
In this paper, we present the effects of dark energy perturbation on the formation and abundance
of cosmic voids.  
We consider dark energy to be a fluid with a negative pressure 
characterised by a constant equation of state $w$ and speed of sound $c_s^2$.
By solving fluid equations for two components, namely, 
dark matter and dark energy fluids, we quantify the effects of dark energy
perturbation on the sizes of top-hat voids. 
We also explore the effects on
the size distribution of voids
based on the excursion set theory.
We confirm that dark energy perturbation negligibly affects 
the size evolution of voids;
$c_s^2=0$ varies the size only by 0.1\% as compared to
 the homogeneous dark energy model. 
We also confirm that dark
energy perturbation suppresses the void size when $w<-1$ and enhances
the void size when $w>-1$ \citep{Basse:2011}.  
In contrast to the negligible impact on the size, we 
find that the size distribution function on scales larger
than 10 ${\rm Mpc}/h$ highly depends on dark energy perturbation; 
compared to 
the homogeneous dark energy model, 
the number of large voids of radius $30 {\rm Mpc}$ 
is 25\% larger for the model with $w=-0.9$ and $c_s^2=0$ 
while they are 20\% less abundant for the model with $w=-1.3$ and $c_s^2=0$.

\end{abstract}

%%%%%%%%%%%%%%%%%%%%%%%%%%%%%%%%%%%%%%%%%%%%%%%%%%%%%%%%%%%%%%%%%%%%%
\section{introduction}
\label{sec:introduction}
%%%%%%%%%%%%%%%%%%%%%%%%%%%%%%%%%%%%%%%%%%%%%%%%%%%%%%%%%%%%%%%%%%%%%
One of the greatest mysteries in cosmology is the 
nature of the energy that
accelerates the Universe, that is, dark energy \citep{Riess:1998, Perlmutter:1999}.
The most popular 
cosmological model, namely, the $\Lambda$CDM model,  
regards the energy source to be a cosmological constant 
corresponding to spatially and temporarily
uniform energy density.  
Although the $\Lambda$CDM model best explains
various observations \citep{Planck2015:cosmology, BOSS:DR9,
  Heymans:2013, Suzuki:2012},
the theoretical origin of this energy is
poorly understood \citep{Weinberg:1989}.

Various models regarding dark energy state that 
the constant energy density over space and time can be relaxed \citep{Tsujikawa:2011}.
If the equation of state of dark energy is not $-1$, 
the energy density of the dark energy varies with time.  
If the speed of sound is equal to that of light, the dark energy
fluid is regarded as spatially 
homogeneous because the Jeans length 
corresponds to the horizon scale.  
Such dynamical but spatially homogeneous
dark energy is usually called quintessence \citep{Zlatev:1999}.  
If the speed of sound of dark energy is smaller than that of light, 
the dark energy is spatially perturbed, 
which is realized in models known as k-essence \citep{Chiba:1998,Armendariz:2000}.

One of the most promising tools to reveal the nature of dark energy
is large-scale structures 
\citep[e.g.][]{Lapparent:1986,2dF,SDSS:technical}.
Traditionally, much attention has been paid to high density
structures such as clusters of galaxies.
Some works have focussed on the impact of dynamical dark energy on
  structure formation
\citep{Wang:1998,Chiba:1998,Abramo:2007,Creminelli:2010,Basse:2011,Heneka:2017}.

In contrast, low density structures called voids \citep{Gregory:1978} 
are also considered to probe large-scale structures.  
Voids occupy  
a large fraction of
volume of the universe \citep{Cautun:2014},  
and also characterise the matter distribution at 
a few tens
Mpc scales \citep{Weygaert:2016}.  
Recently, various properties of voids have been systematically studied
\citep{Pan:2012,Sutter:2012a, Cautun:2014}.
Voids provide us with independent tools for probing the cosmological
models in different manners from overdense objects.
Although the voids are described in the quasi linear regime and thus
  are expected to be more robust probes for large scale structures \citep{Weygaert:2016},
  defining the voids from data remains ambiguous.
  Therefore, testing the cosmological model by
  combining different observables where systematics and cosmological
  sensitivity are different is essentially important.
  
Voids are used for constraining cosmological models mainly in two ways so far.  
Firstly, the shape of the void which is sensitive 
the dark energy models can be potentially used to constrain cosmology 
 \citep{Park:2007,Lee:2009,Biswas:2010,Bos:2012}.
The shape of the void is also useful in the context of 
the Alcock Paczynski (AP) test \citep{Alcock:1979}.
The AP test was first applied to the separation of 
quasar pairs \citep{Phillipps:1994}, 
and \cite{Hu:2003}
proposed the use of the Baryon Acrostic Oscillation (BAO) ring as a more promising probe.
Voids can be also used as the probes for the AP test because 
the cosmological principle ensures that the voids are expected
to be spherical on average \citep{Ryden:1995,Ryden:1996,Lavaux:2012}. 

On the contraly, it is known that the shape of void is correlated with
each other up to scales more than 30 Mpc \citep{Platen:2008}, and thus
one needs to take into account such an alignment of voids to avoid a
systematic effect when applying voids to the AP test.

\cite{Sutter:2012} first applied the AP test to stacked voids 
but found no significant signals because the number of voids was insufficient.
 \cite{Sutter:2014} revisited the analysis using the Sloan Digital Sky Survey (SDSS)
Data Release 10 \citep{SDSS:DR10}
and found a substantial signal for the AP test.
Recently, \cite{Mao:2017} have put a constraint on $\Omega_{m}$ 
using the AP test on the SDSS Data Release 12 \citep{SDSS:DR12}.

Secondly, voids are used for constraining cosmological models 
based on their abundance.  
This statistics is also known to be sensitive to dark energy models \citep{Pisani:2015}.  
The theoretical prediction for void abundance is 
given by \cite{Sheth:2004} (the SVdW model).
They adopted the excursion set theory \citep{Bond:1991} to predict  
the mass fraction involved in the void region.
They assumed that the abundance of 
spherically symmetric
isolated voids arises when the density contrast becomes 
less than $\delta_m \simeq -0.8$ \citep{Blumenthal:1992}.
The original SVdW model is 
inconsistent with both the abundance obtained from N-body
simulations and real galaxy distributions.  
Previous works have extended the SVdW model 
by relaxing the constant density threshold for void formation, 
which is calibrated via N-body simulations
\citep{Achitouv:2015},
or by considering the formation threshold as a free parameter
\citep{Chan:2014,Nadathur:2015}
in order to obtain a good agreement of the model with the N-body simulations.
However,  the best-fit values of the formation threshold are not 
physically interpreted.

With regard to the cosmological constrains using voids, 
the parameters of the equation of state of dark
energy have been well studied in the literature \citep{Pisani:2015} 
but the inhomogeneity of the dark energy model 
where dark energy is spatially perturbed has not been focused on very much.
\cite{Novosyadlyj:2016} studied
the effect of dark energy perturbation
on the density and velocity profiles of voids.
They adopted the universal density profile 
\citep{Hamaus:2014a}
and found that dark energy perturbation 
negligibly affects the profiles.

In this paper, we investigate the impacts
of dark energy perturbation on 
the evolution of isolated top-hat spherically symmetric voids, 
focusing on their size and formation epoch.
We also study the mechanism by which dark energy perturbation influence void abundance 
by adopting the SVdW model.
We find that the void abundance  
is remarkably sensitive
to dark energy perturbation.

This paper is organised as follows.
In Section \ref{sec:model}, following the discussion of \cite{Basse:2011}, we revisit the
spherical collapse model under dark energy perturbation 
and apply it to the spherical void formation.
Section \ref{sec:results_size} is devoted to the numerical
  results of the void size evolution.
In Section \ref{sec:voidsize}, we present the dependence of void
  abundance on dark energy perturbation.
We provide a summary in Section \ref{sec:summary}.

Throughout this work, we assume the cosmological parameters as
$h=0.7, \Omega_{m,0} = 0.3, $ and $\Omega_{Q,0} = 0.7$,
which are dimensionless Hubble parameter, density parameters of matter and dark energy, respectively,
where we denote the matter component as $m$
the dark energy component as $Q$. 
 In addition, we assume that parameters of the equation of state and the speed of sound
  are constant, and we take the natural unit of $c=1$.

%%%%%%%%%%%%%%%%%%%%%%%%%%%%%%%%%%%%%%%%%%%%%%%%%%%%%%%%%%%%%%%%%%%%%
\section{Spherical void formation in the inhomogeneous dark energy model}
\label{sec:model}
%%%%%%%%%%%%%%%%%%%%%%%%%%%%%%%%%%%%%%%%%%%%%%%%%%%%%%%%%%%%%%%%%%%%%

Void is a structure that on average has an under density profile compared
  to the mean density of the universe. The density in the centric
  regions are in general less dense compared to the outer regions. 
 Such a density gradient 
makes the inner shell of voids expand faster than outer shells. 
  Then matter in a void region accumulates around the void to form dense
  ridges as their boundaries and such a process results in forming a density profile like a top-hat shape
  even if we assume a smoothed initial density profiles \citep{Sheth:2004}.

On the other hand, \cite{ Hamaus:2014} has reported that the spherically averaged 
density profile of voids in N-body simulations is less steep than expected from the top-hat shape.
The difference comes from the fact that individual voids in the realistic situations are not spherical 
because individual voids are not isolated and their dynamical evolution  
is highly affected by their local environment.
Although an analytic calculation was shown that an isolated void tends to
be spherical as it evolves \citep[e.g.][]{Icke:1984},
individual voids 
do not maintain the shapes at the initial condition and their shapes are far from the
spherical ones \citep{Platen:2008}.
However, if one defines the radial profile by measuring the density on the equidistance contour from the boundary of the void, 
the stacked profile resembles compensated top-hat shape \citep{Cautun:2016}.
Therefore, the top-hat profile seems to be a generic property of voids 
even in the realistic situations \citep{Dubinski:1993,Weygaert:1993,Sheth:2004,Weygaert:2016}

The moment when the inner shells catch up with the outer shells is defined as 
{\it shell crossing}  
and previous studies have pointed out 
that the evolution of voids is in full non-linear regime at the moment
  of shell crossing
\citep{Suto:1984,Fillmore:1984,Bertschinger:1984}.  
The shell crossing
for an isolated spherically symmetric void with 
a top-hat density profile in the Einstein-de Sitter universe 
occurs when the density contrast reaches $\delta^{\rm TH}_m \simeq
-0.8$ \citep{Blumenthal:1992, Dubinski:1993}. 

Even though a simple model which assumes isolated spherical void with a top-hat profile 
does not perfectly 
describe the actual dynamical evolution of voids in the complex environment,
starting with such a simplified model can
provide us with a lot of insights on the effects of dark
  energy model. Considering more realistic situations is beyond the
scope of this paper and we will explore more detailed study with
numerical experiments as a future work.
In this section, we describe the formation of a spherically symmetric isolated void 
with a top-hat density profile in the presence of dark energy perturbation.  

In Section \ref{ssec:spherical}, we revisit the spherically
symmetric formation model following the discussion of \cite{Basse:2011}.
In Section \ref{ssec:depert}, we extend the model in the presence of
dark energy perturbation for constant speed of sound
and constant equation of state parameters.

%----------------------------------------------------------------------------
\subsection{Spherical symmetric model}
\label{ssec:spherical}
%----------------------------------------------------------------------------
To trace the non-linear evolution of voids, we consider a spherically
symmetric underdense region with a top-hat profile of radius $R_{\rm v}$.  The
evolution of 
the void in an expanding universe follows the equation of motion

\begin{equation}
	\frac{1}{R_{\rm v}}\frac{d^2R_{\rm v}}{dt^2}=-\frac{4\pi G}{3} \sum_n (\rho_n + 3P_n),
	\label{eq:spherical_phys}
\end{equation}

where  
$\rho_n$ and $P_n$ are, respectively, the energy density and pressure of the 
fluid component $n$ enclosed in the region $r \leq R_{\rm v}$. 
The density $\rho_n$ and pressure $P_n$ can be decomposed into
unperturbed and perturbed components as
\begin{equation}
	\rho_n =\bar{\rho}_n +\bar{\rho} _n \delta_n,
	\label{eq:density}
\end{equation}
\begin{equation}
	P_n = \bar{P}_n + \delta P_n.
	\label{eq:pressure}
\end{equation}
where $\bar{\rho}_n$ and $\bar{P}_n$ are the mean background values.
Since we assume the top-hat density profile, 
the density contrast can be expressed as
\begin{align}
	\delta_n(t,r) = 
		\begin{cases}
			\ \delta^{\rm TH}_n(t) &(0 \leq r \leq R_{\rm v}),\\
			\ 0		&(R_{\rm v} < r).
		\end{cases}
	\label{eq:profile}
\end{align}
Given that the void evolves in an expanding universe, equation
  (\ref{eq:spherical_phys}) can be rewritten in terms of the comoving coordinate as
\begin{equation}
  	\frac{\ddot{X}_{\rm v}}{X_{\rm v}} + {\cal H}\frac{\dot{X}_{\rm v}}{X_{\rm v}} = -\frac{4\pi G}{3} a^2 \sum_n ( \bar{\rho}_n \delta^{\rm TH}_n+ 3 \delta P^{\rm TH}_n),
  	\label{eq:spherical_com}
\end{equation}
where $X_{\rm v}$, $a$ and ${\cal H}$ represent the comoving radius of the void,
scale factor, and conformal Hubble parameter, respectively.
The dot denotes the derivative with respect to the conformal time, $d\tau = dt/a$.

%----------------------------------------------------------------------------
\subsection{Dark energy perturbation}
\label{ssec:depert}
%----------------------------------------------------------------------------
Now we assume that the universe consists of only dark matter
  and dark energy.
Here we consider the model wherein dark energy and matter are
  spatially perturbed.
Therefore, equation (\ref{eq:spherical_com}) can be written as
  \begin{align}
   	 \frac{\ddot{X}_{\rm v}}{X_{\rm v}} + {\cal H}\frac{\dot{X}_{\rm v}}{X_{\rm v}} = -\frac{4\pi G}{3} a^2[ \bar{\rho}_m \delta^{\rm TH}_m 
  	+ \bar{\rho}_Q\delta^{\rm TH}_Q + 3 \delta P^{\rm TH}_Q].
	\label{eq:spherical_com_mde}
\end{align}

In order to solve equation (\ref{eq:spherical_com_mde}), 
we must  simultaneously solve the evolution of matter and dark energy 
perturbations. 
Given that the mass of the matter inside a void 
is conserved,
\begin{equation}
	M=\frac{4 \pi}{3} \bar{\rho}_m(1+\delta_m^{\rm TH}) R_{\rm v}^3  = \frac{4 \pi}{3} \bar{\rho}_{m}(1+\delta_{m}^{\rm TH}) a^3 X_{\rm v}^3.
	\label{eq:mass_conservation}
\end{equation} 
The matter density perturbation at a given time $\tau$ is 
simply scaled by the void radius as
\begin{align}
	\delta_m^{\rm TH}(\tau) = (1 + \delta^{\rm TH}_{m,i}) \left[ \frac{X_{{\rm v},i}}{X_{\rm v}(\tau)} \right]^3 -1,
	\label{eq:mass_conservation2}
\end{align}
where the variable with subscript $i$ takes the value at the initial
time $\tau_i$.

In order to introduce the effect of dark energy pressure into
a Newtonian regime, we adopt a pseudo Newtonian
approach \citep{Lima:1997,Basse:2011}. 
The continuity and Euler equations are given as
\begin{align}
	& \dot{\rho}_Q+3 {\cal H} (\rho_Q +P_Q) +  \nabla \cdot [(\rho_Q + P_Q) \bm{v}_Q]=0,
	\label{eq:continuous}\\
	& \dot{{\bm v}}_Q+{\cal H} {\bm v}_Q+({\bm v}_Q \cdot \nabla) {\bm v}_Q + 
	\frac{\nabla P_Q + {\bm v}_Q \dot{P}_Q}{\rho_Q+P_Q} + \nabla \phi = 0,
	\label{eq:Euler}
\end{align}
where $\bm{v}_Q$ is the velocity of dark energy fluid, 
and $\phi$ is a gravitational potential that satisfies the Poisson equation,
 \begin{equation}
  	\nabla^2 \phi 
  	= 
  	4\pi G a^2 
  	[ 
  	\bar{\rho}_m \delta_m + \bar{\rho}_Q \delta_Q + 3 \delta P_Q
  	].
\label{eq:Poisson}
\end{equation}
Substituting equations (\ref{eq:density}) and (\ref{eq:pressure}) into (\ref{eq:continuous})
and (\ref{eq:Euler}) yields the evolution equations for dark energy perturbation;
\begin{align}
	&\dot{\delta}_Q+3 {\cal H}(\delta_Q^P - w \delta_Q)
	+\nabla \cdot [(\rho_Q + P_Q) {\bm v}_Q / \bar{\rho}_Q]=0, 
	\label{eq:continuous_delta}\\
	& \dot{{\bm v}}_Q + {\cal H}{\bm v}_Q + ({\bm v}_Q \cdot \nabla) {\bm v}_Q +
	\frac{\nabla \delta_Q^P + {\bm v}_Q (w \dot{\bar{\rho}}_Q /\bar{\rho}_Q + \dot{\delta}_Q^P)}{1+w+\delta_Q+\delta_Q^P}
	+\nabla \phi = 0,
	\label{eq:Euler_delta}
\end{align}
where $w=\bar{P}_Q/\bar{\rho}_Q$ is the equation of state 
of the dark energy component, 
and $\delta_Q^P=\delta P_Q / \bar{\rho}_Q$. 
We assume that dark energy perturbation is always subdominant compared to 
matter perturbation.  Thus equations (\ref{eq:continuous_delta}) and (\ref{eq:Euler_delta}) are
simplified in the linear perturbation regime as,
\begin{align}
\label{eq:continuous_lin}
& \dot{\delta}_Q ^{\rm lin} + 3 {\cal H} (\delta_Q^{\rm P, lin} -w \delta_Q^{\rm lin}) + (1+w) \theta _Q ^{\rm lin}=0, \\
\label{eq:Euler_lin}
& \dot{\theta}_Q ^{\rm lin}+ (1-3w) {\cal H} \theta_Q ^{\rm lin}+ \frac{\nabla^2 \delta_Q^{\rm P, lin}}{1+w} + \nabla ^2 \phi = 0,
\end{align}
where we define the divergence of velocity, 
$\theta _Q= \nabla \cdot {\bm v}_Q$. 
As we will see in Section \ref{sec:results_size}, the linear regime
assumptions for dark energy perturbation are well satisfied.
The above equations can be more easily handled in Fourier space.
Since we do not assume the adiabatic condition for dark energy, its pressure perturbation  
 is expressed by density and entropy perturbations. 
In Fourier space and in the conformal Newtonian gauge, the pressure perturbation for
  dark energy can be given as \citep{Bean:2004}
\begin{align}
	 \tilde{\delta}_Q^{\rm P, lin} = c_s^2 \tilde{\delta}^{\rm lin}_{Q}+ 3
			{\cal H}(1+w)(c_s^2 -w) \tilde{\theta}_Q^{\rm lin}/k^2,
	\label{eq:deltap_gi}
\end{align}
where the speed of sound of dark energy $c_s^2$ is defined in the dark
energy rest frame as
\begin{align}
	c_s^2 = \left. \frac{\delta P_Q}{\bar{\rho}_Q \delta_Q} \right|_{\rm rest}.
	\label{eq:sound_speed}
\end{align}
The speed of sound of dark energy characterises the typical scale of
dark energy perturbation and is assumed to be constant in this work.
Owing to the second term in equation (\ref{eq:deltap_gi}), the pressure perturbation of dark energy 
does not vanish even when $c_s^2=0$.

\begin{figure}
  \includegraphics[width=1.0 \linewidth]{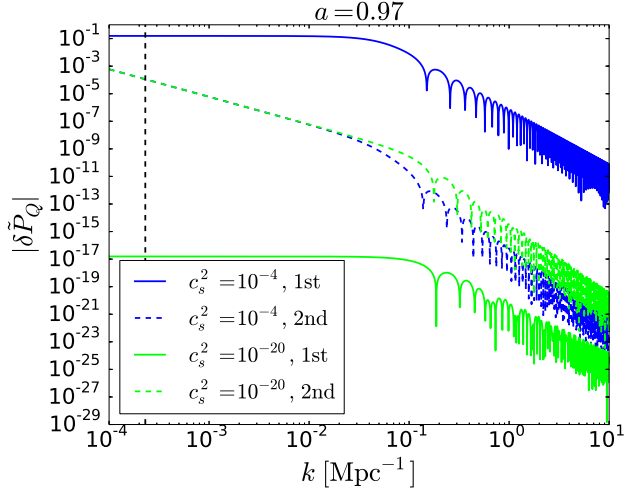}
  \caption{ 
    Contributions of the first and second terms 
    to $\tilde{\delta}_Q^{\rm P, lin}$
    in equation (\ref{eq:deltap_gi}) at $a=0.97$.
    The solid and dashed lines denote the individual contributions of
      the first and second terms, and the green and blue lines correspond to
      $c_s^2=10^{-4}$ and $c_s^2=10^{-20}$, respectively.
    For $c_s^2=10^{-4}$, the second term is smaller than the first
    term by the factor of $10^{-3}$ even at the horizon scale of the
      universe which is indicated
    by the black vertical dashed line.  For $c_s^2 =10^{-20}$, the
    second term 
    dominates the first term over all scales.
    We use the initial conditions presented in Section \ref{ssec:initial_conditions},
    but set $w = -0.9$ and $M = 10^{15} M_{\sun}$.
    The result is insensitive to the choice of $w$ but the oscillatory damping scale depends on 
     the void mass.    
    \label{fig:delta_pk}}
\end{figure}

In Fig. \ref{fig:delta_pk}, we see that for $c_s^2=10^{-4}$ the amplitude of the second term is smaller 
than that of the first term by the factor of at least $10^{-3}$ 
over all scales inside the horizon,
whereas for $c_s^2=10^{-20}$ the second term 
dominates over the first term.
We note that even for scales well inside the horizon scale, 
pressure perturbation of dark energy due to entropy perturbation
is not negligible for $c_s^2\ll 1$.

By substituting equation (\ref{eq:deltap_gi}) into equation (\ref{eq:Euler_lin})  in Fourier space, 
we obtain 
\begin{align} 
	&\dot{ \tilde{\delta} }_Q ^{\rm lin} + 3 (c_s^2 - w ) {\cal H} \tilde{\delta}_Q^{\rm lin} + (1 + w ) \tilde{\theta} ^{\rm lin}= 0,
	\label{eq:continuous_delta_lin_k}\\
	&\dot{ \tilde{\theta} }_Q ^{\rm lin}+ (1- 3 c_s^2) {\cal H}  \tilde{\theta}^{\rm lin} - \frac{k^2 c_s^2}{1 + w} \tilde{ \delta}_Q ^{\rm lin} -k^2 \tilde{\phi} =0.
	\label{eq:Euler_delta_lin_k}
\end{align}
We use the Poisson equation in Fourier space along with the above equations,
\begin{align}
	-k^2 \tilde{\phi} = 4\pi G a^2 
  	[ \bar{\rho}_m \tilde{\delta}_m + \bar{\rho}_Q \tilde{\delta}_Q + 3 \tilde{\delta P}_Q ]. 
  \label{eq:poisson_fspace}
\end{align}
As can be clearly seen, the dark energy perturbation affects 
structure formation only thorough the gravitational force through the
Poisson equation.

For later convenience, we explicitly give the linearized equations for
matter perturbation as follows; 
\begin{align}
	&\dot{\tilde{\delta}}_m^{{\rm lin}} + \tilde{\theta}_m^{{\rm lin}} =0,
	\label{eq:continuous_deltam_lin_k}\\
	&\dot{\tilde{\theta}}_m^{{\rm lin}} +{\cal H} \tilde{\theta}_m^{{\rm lin}} 
	+ 4\pi G a^2 [ \bar{\rho}_m \tilde{\delta}_m^{{\rm lin}} +\bar{\rho}_Q\tilde{\delta}^{\rm lin}_Q + 3 \tilde{\delta P}^{\rm lin}_Q ] =0.	
	\label{eq:Euler_deltam_lin_k}
\end{align}

From equations (\ref{eq:continuous_delta_lin_k}),
(\ref{eq:Euler_delta_lin_k}), and (\ref{eq:poisson_fspace}),
eliminating $\theta$ yields the evolution equation of the linear dark
energy perturbation;

\begin{align}
	&\frac{ d ^2 \tilde{\delta}^{\rm lin}_{Q}}{d s^2} + {\cal D}(s) \frac{ d \tilde{\delta}^{\rm lin}_{Q}}{d s}
	+ \left[ \frac{k^2 c_s^2}{{\cal H}^2} \chi (s) - \kappa(s) \right] \tilde{\delta}^{\rm lin}_{Q}\nonumber\\
	&\hspace{3cm}= \frac{3}{2} (1+w)\Omega_m(s) \tilde{\delta}_m,
	\label{eq:2nd_deltade_k}
\end{align}
where
\begin{align}
	&s = \ln a,\\
	&{\cal D}(s) \equiv 1 + \frac{1}{\cal H} \frac{d {\cal H}}{ds} - 3w,\\
	&\kappa(s) \equiv 3w \left( 1+ \frac{1}{\cal H} \frac{d {\cal H}}{ds} \right)+\frac{3}{2} (1+w) \Omega_{Q}(s),\\
	&\chi (s) \equiv 1+3 \frac{{\cal H}^2}{k^2} \left[ 1+ \frac{1}{\cal H} \frac{d {\cal H}}{ds}  - 3(c_s^2 -w)\right. \nonumber\\
	&\hspace{4cm} \left. - \frac{3}{2} (1+w) \Omega_{Q}(s) \right] .
\end{align}
The evolution equation is numerically solved once the
  initial condition of the dark energy and matter perturbations are specified as described in the
  next section.
  \begin{figure*}
  \begin{tabular}{cc}
    \includegraphics[width=0.5\linewidth]{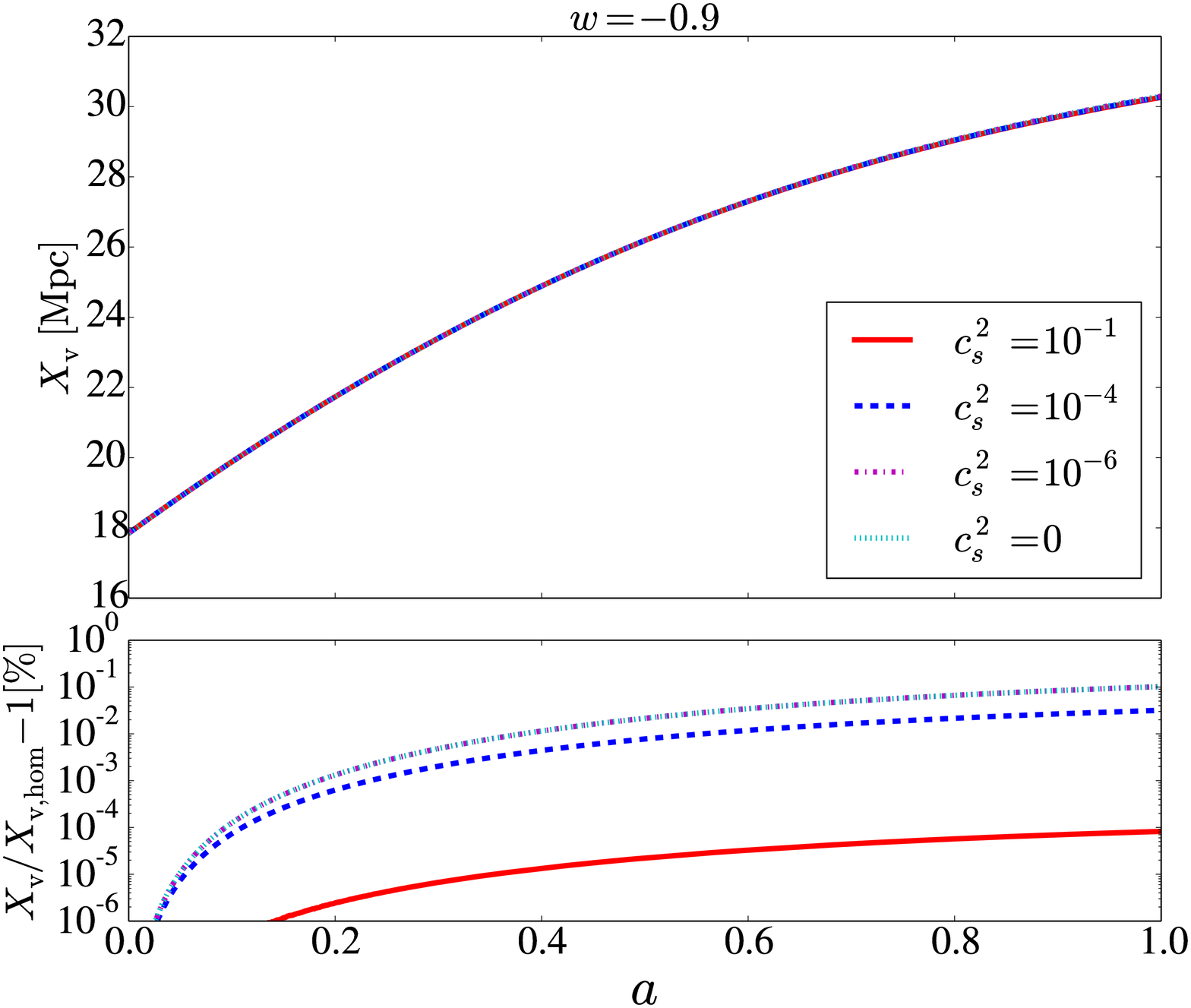} &
    \includegraphics[width=0.5\linewidth]{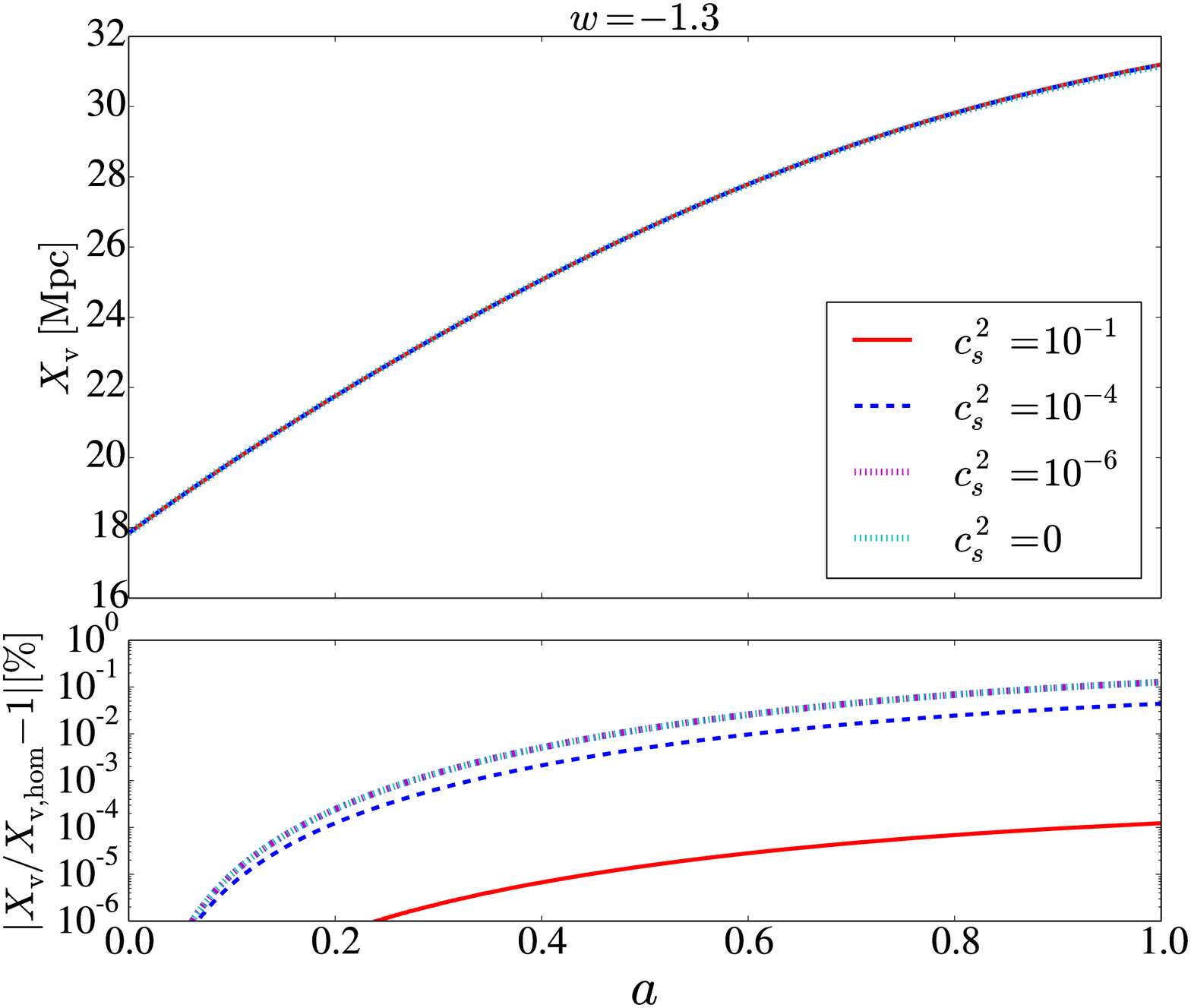}
  \end{tabular}
 \caption{ Size evolutions of voids with different speed of sound.  
 		We set $w=-0.9$ for the left panels and $w=-1.3$ for the right panels.  
		The speed of sound are $c_s^2=0,\ 10^{-6},\ 10^{-4}$, and $10^{-1}$ both for the left and right panels.  
		 We compare the size evolutions for each value of the speed of sound 
		 with the homogeneous
                 dark energy model in the lower panels.  
		 The absolute value of fractional difference for the lower right panel corresponds to 
		 that the dark energy clustering for $w<-1$ 
		 suppresses the void evolution and the difference takes negative values.	 
		 \label{fig:radius_cs}}
\end{figure*}

 %----------------------------------------------------------------------------
\subsection{Initial conditions}
\label{ssec:initial_conditions}
%----------------------------------------------------------------------------
Now we set up the initial conditions for the calculations.
\begin{itemize}
\item{Initial time}\\
Following \cite{Basse:2011}, we begin our calculation at the
  cosmic time
  \begin{align}
    t_iH_0 = 2.0 \times 10^{-6},
  \end{align}
which corresponds to the scale factor,
  \begin{align}
    a_i = a_0 \left( \frac{3 t_i H_0 \sqrt{\Omega_{m,0}}}{2} \right) ^{2/3} \approx 1.4 \times 10^{-4} .
  \end{align}
\item{Initial matter density contrast}\\
  In the EdS universe, linear matter fluctuation evolves
  in proportion to the scale factor.
   Thus, at arbitrary time, it scales as
  \begin{align}
    \delta_m^{\rm lin}(a) = \frac{a(\tau)}{a_v} \delta_v,
  \end{align}
  where $\delta_v$ is the critical density for void formation and $a_v$ is 
  the scale factor when the linear density contrast reaches $\delta_v$.  
  For the EdS universe, 
  we analytically obtain $\delta_v=-2.717$ \citep{Blumenthal:1992}.
  If $\delta_m^{\rm lin}$ reaches $-2.717$ at the present time, 
  the corresponding initial matter density contrast
  is $\delta_{m,i}^{\rm lin} \simeq \delta^{\rm TH}_{m,i} = -3.8 \times 10^{-4}$
  in the EdS universe. 
  For the universe with dark energy, matter growth is 
  suppressed at a later epoch compared to that in the EdS universe.  
  To allow voids to form by the present time, 
  we give a slightly larger negative value of matter density contrast than that in the EdS universe,
  \begin{align}
    \delta_{m,i}^{\rm lin} = \delta^{\rm TH}_{m,i} = -5.0 \times 10^{-4},
  \end{align}
  where we assume that in the very early epoch the density fluctuation can be well described by the
  linear perturbation.  
Note that the choice of the values of initial matter density contrast does
  not affect the comparison of void size among different dark energy models.
    
\item{Initial dark energy density contrast}\\
  We assume that the density contrast of dark energy is proportional to 
  that of matter.  Thus once the initial condition of matter is specified, the initial
  condition of the dark energy perturbation is immediately obtained by
  numerically integrating \rref{eq:2nd_deltade_k}.

\item{Initial velocity}\\
  We obtain the initial velocity of a void shell by differentiating equation (\ref{eq:mass_conservation}),
  \begin{align}
    \left. \frac{dX}{d\tau}\right|_{\tau=\tau_i} = \frac{2 a_i}{3\eta_i} \left(1- \frac{1}{3}\delta^{\rm TH}_{m}(\tau_i) \right)
    - {\cal H}_{i},
  \end{align}
where $\eta_i = \int ^{\tau _i}_{0} a d\tau$.
\end{itemize}

 %----------------------------------------------------------------------------
\subsection{Summary for calculus}
\label{ssec:Mini_summary}
%----------------------------------------------------------------------------

To clarify our procedures, here we provide a summary
  of our calculation steps.

 \begin{enumerate}
 \item{}
   We set the initial matter fluctuation and the mass of matter inside to derive
   the radius of the top-hat initial void by using equation (\ref{eq:mass_conservation}).
   
   \item{}
  We conduct Fourier transform of $\delta_m$ to find
  $\tilde{\delta}_m$ as 
  \begin{align}
    \tilde{\delta}_m(k,\tau) &= \int ^{X_{\rm v}}_0 d^3 x\  \delta_{m}^{\rm TH}(\tau)\  e^{-i \bm{k}\cdot \bm{x}}\nonumber\\
    					& = \frac{4 \pi}{3} \delta_m^{\rm TH} (\tau)X_{\rm v}^3 (\tau)\tilde{W}(kX_{\rm v}),
  \end{align}
  where $\tilde{W}(kX_{\rm v})$ is a top-hat window function in Fourier space,
  \begin{align}
  	 \tilde{W}(kX_{\rm v}) = \frac{3\left[ \sin(kX_{\rm v}) - kX_{\rm v} \cdot \cos(kX_{\rm v}) \right]}{(kX_{\rm v})^3}.
\end{align}
  Then we solve the differential equation (\ref{eq:2nd_deltade_k}) for
  $\tilde{\delta}_Q^{\rm lin}$.
\item{}
  We conduct Fourier inverse transform of $\tilde{\delta}_Q^{\rm lin}$ to
  obtain $\delta^{\rm TH}_Q$ in real space, which is averaged out
  inside the void radius,
  \begin{align} 
  	\delta^{\rm TH}_Q (\tau)= \frac{3}{4\pi X_{\rm v}^3} \int d^3 x W(|\bm{x}|) \delta^{\rm lin}_Q(\bm{x},\tau).
  \end{align}
   Thus the relation between $\delta_Q^{\rm TH}$ and $\tilde{\delta}_Q^{\rm lin}$ is
\begin{align}
	\delta^{\rm TH}_{Q}(\tau) = \int \frac{dk}{2\pi^2} k^2 \tilde{W}(kX_{\rm v}(\tau))\tilde{\delta}^{\rm lin}_{Q}(k,\tau).
	\label{}
\end{align}

\item{}
We solve the second order differential  equation for $X_{\rm v}(\tau)$ using
the obtained $\delta ^{\rm TH}_m$ and $\delta_Q^{\rm TH}$.
\item{}
Next, we calculate $\delta^{\rm TH}_m$ using \rref{eq:mass_conservation2}
\item{}
For the next time step, we repeat steps (ii) to (v) untill the matter density
fluctuation reaches $\delta^{\rm TH}_m=-0.8$.
\end{enumerate}

%----------------------------------------------------------------------------
\section{Results of size evolution}
\label{sec:results_size}
%----------------------------------------------------------------------------

In this section, we present the numerical results for the evolution of a spherically
symmetric void with a top-hat density profile.
We 
explore the dark energy parameters: 
$0\leq c_s^2 \leq 1$, $w=-0.9$ and $-1.3$.
We also examine the dependence of the void evolution on the void mass.
We set the initial mass 
of $10^{13}M_{\sun}$,  $10^{15}M_{\sun}$,  and
$10^{17}M_{\sun}$.
The void mass only affects the void size through equation
  (\ref{eq:mass_conservation}) and the corresponding initial radii are 
$0.83, 18$, and $83{\rm Mpc}$, respectively, in the comoving scale.

\subsection{Dependence on $c_s^2$}
  In Fig. \ref{fig:radius_cs},
  we show the dependence of 
  the evolution of voids on the speed of sound.  We set $w=-0.9$ for the left
  panel, $w=-1.3$ for the right panel and $M_i = 10^{15}M_{\sun}$ for
  both panels.  The upper panels show the evolution of void radii; the
  lower panels show the fractional differences in void radii with
  respect to that 
  for the homogeneous dark energy model ($c_s^2 = 1$).
  We see that if the speed of sound is low, the deviation is large.  
  For $w=-0.9$ the small speed of sound enhances the size of the
  void whereas for $w=-1.3$, the small speed of sound suppresses the evolution of voids.    
  We can not find any differences in void size 
  between $c_s^2 =0$ and $c_s^2=10^{-6}$  
  for both $w=-0.9$ and $-1.3$ because the void size is larger than the 
    Jeans length of the fluctuations even for $c_s^2=10^{-6}$.

For both $w=-0.9$ and $-1.3$, the difference in void size
  between the $c_s^2=0$ model and the homogeneous dark energy
  model is of the order of 0.1\%.
  For $c_s^2 =10^{-1}$, the difference is $10^{-4}$\% both for $w=-0.9$ and $-1.3$, 
  which is much smaller than that for $c_s^2=0$. 
  This is because the Jeans length for $c_s^2=10^{-1}$ is always  larger 
  than the void size during evolution.  
  Therefore the effect of enhancement or suppression due to dark energy clustering is negligible.
  
\subsection{Dependence on $w$}
  In Fig. \ref{fig:radius_cs}, we see that dark energy perturbation acts in opposite ways
   depending on the value of $w$:
   it suppress the evolution of voids for $w<-1$ while it enhances for $w>-1$.
   
  In order to understand these opposite effects, we consider the
  equation for the evolution of the dark energy perturbation 
  (\rref{eq:2nd_deltade_k}).  
  If $w>-1$, the density contrast of the dark energy component 
  evolves with the same sign as the matter density contrast 
  whereas if $w<-1$, the sign is opposite to the matter component.  
  Then, in equation (\ref{eq:spherical_com}), if the density
  contrast of dark energy has the same sign as that of matter, the 
  fluctuation of gravitational potential is enhanced, 
  promoting the growth of the voids.  
  Conversely, if the 
  dark energy density contrast has an opposite sign
  as that of matter,
  the gravitational potential fluctuation is reduced,
  suppressing the evolution of void size.
  
  We also see that for $w =-0.9$, dark energy perturbation affects
  relatively earlier than for $w = -1.3$.  
  This can be simply 
  explained by the domination epoch of dark energy in the cosmic time.
  In the universe where $w<-1$, dark energy dominates the expansion
  at a later epoch and therefore, there is less chance that the matter
  fluctuations are affected by dark energy. 

\subsection{Dependence on initial mass}
\begin{figure}
  \includegraphics[width=1 \linewidth]{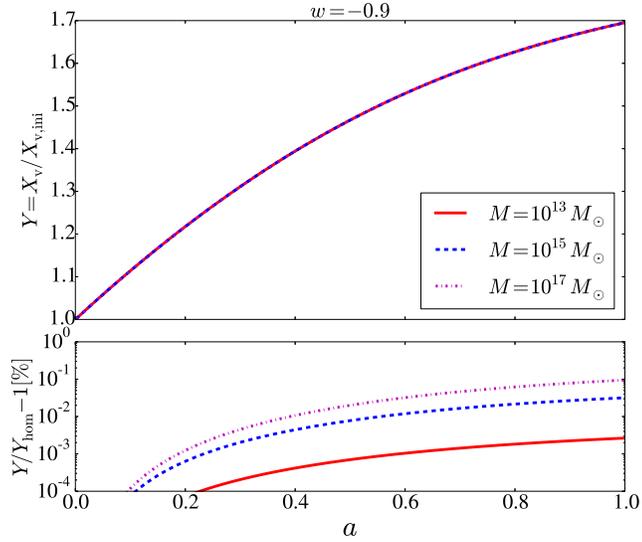}
  \caption{ 
    Same as Fig. \ref{fig:radius_cs} for $w=-0.9$ and $c_s^2=10^{-4}$ but for
    an initial void mass of $10^{13}, 10^{15}$ and $10^{17}M_{\sun}$.
    The upper panel displays the void sizes normalised by the initial size.
        The lower panel shows fractional differences of the void sizes 
    normalised by that in the homogenous dark energy model.
    \label{fig:radius_M}}
\end{figure}

\begin{figure}
	 \includegraphics[width=1\linewidth]{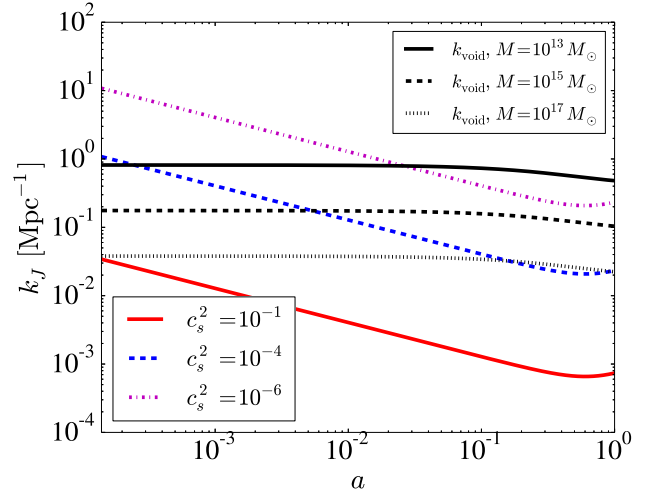} 
	\caption{
	Evolution of the Jeans wavenumbers for each value of speed of sound (coloured lines)
     	and the wavenumber which corresponds to the void radius for each mass scale with $c_s^2=10^{-1}$(black lines)
	as indicated in the figure. }
	\label{fig:Jeans_wave}
\end{figure}

In Fig. \ref{fig:radius_M}, we plot the
evolution of void size with different void masses.
We set $w=-0.9$ and $c_s^2=10^{-4}$
but take values for
the initial masses
of $10^{13}$, $10^{15}$ and $10^{17}M_{\sun}$.
Here we choose $c_s^2=10^{-4}$ in order to demonstrate the relation 
between the void size and the Jeans length.
As we see in Fig. \ref{fig:Jeans_wave}, the Jeans length for $c_s^2 =10^{-6}$
is always smaller than the void size for 
$M>10^{15}M_{\sun}$.
Therefore we do not find any difference in the evolution of void size.

In Fig. \ref{fig:radius_M}, we see
 that the void with larger initial mass is more affected 
by the dark energy perturbation.
  As we see in the next section, if
  the size of a void exceeds the Jeans length of dark energy, 
  the density contrast of dark energy inside the void evolves 
  regardless of the size of the void.
  However, if the size is less than the Jeans length, 
  the evolution of the dark energy perturbation inside the void 
  is suppressed.

\subsection{Evolution of dark energy perturbation}
\begin{figure*}
  \begin{tabular}{ccc}
    \includegraphics[width=0.33\linewidth]{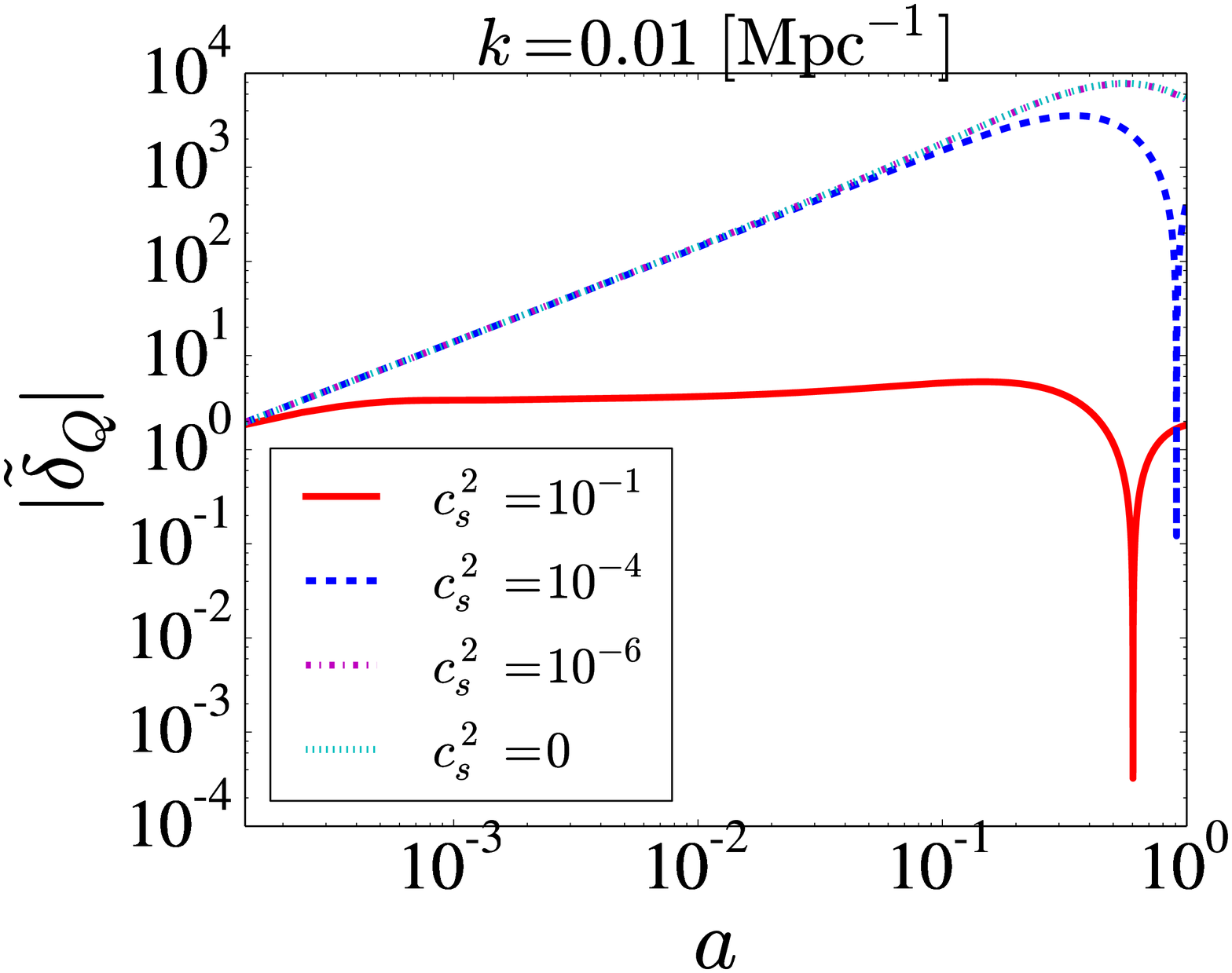}
    \includegraphics[width=0.33\linewidth]{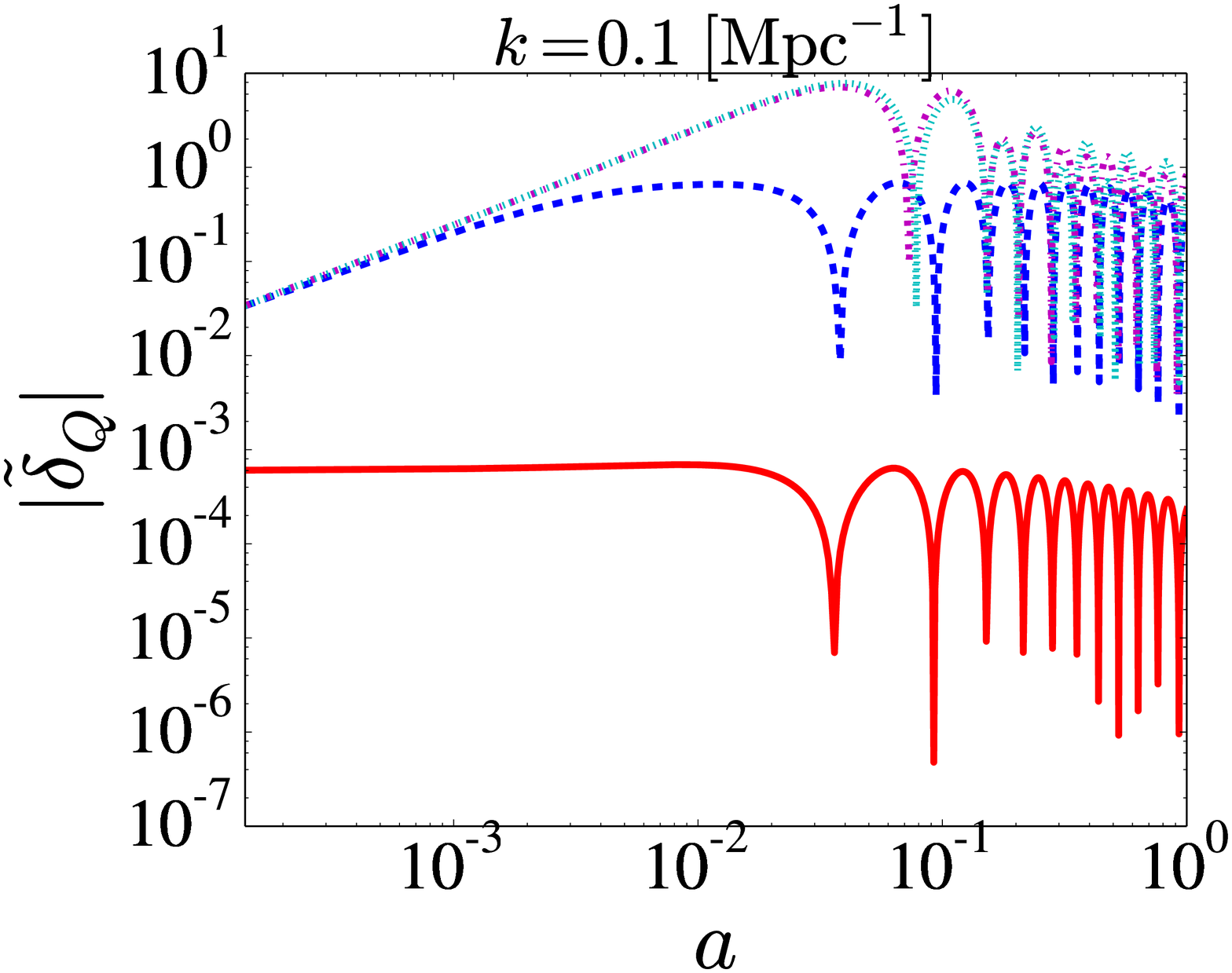} 
    \includegraphics[width=0.33\linewidth]{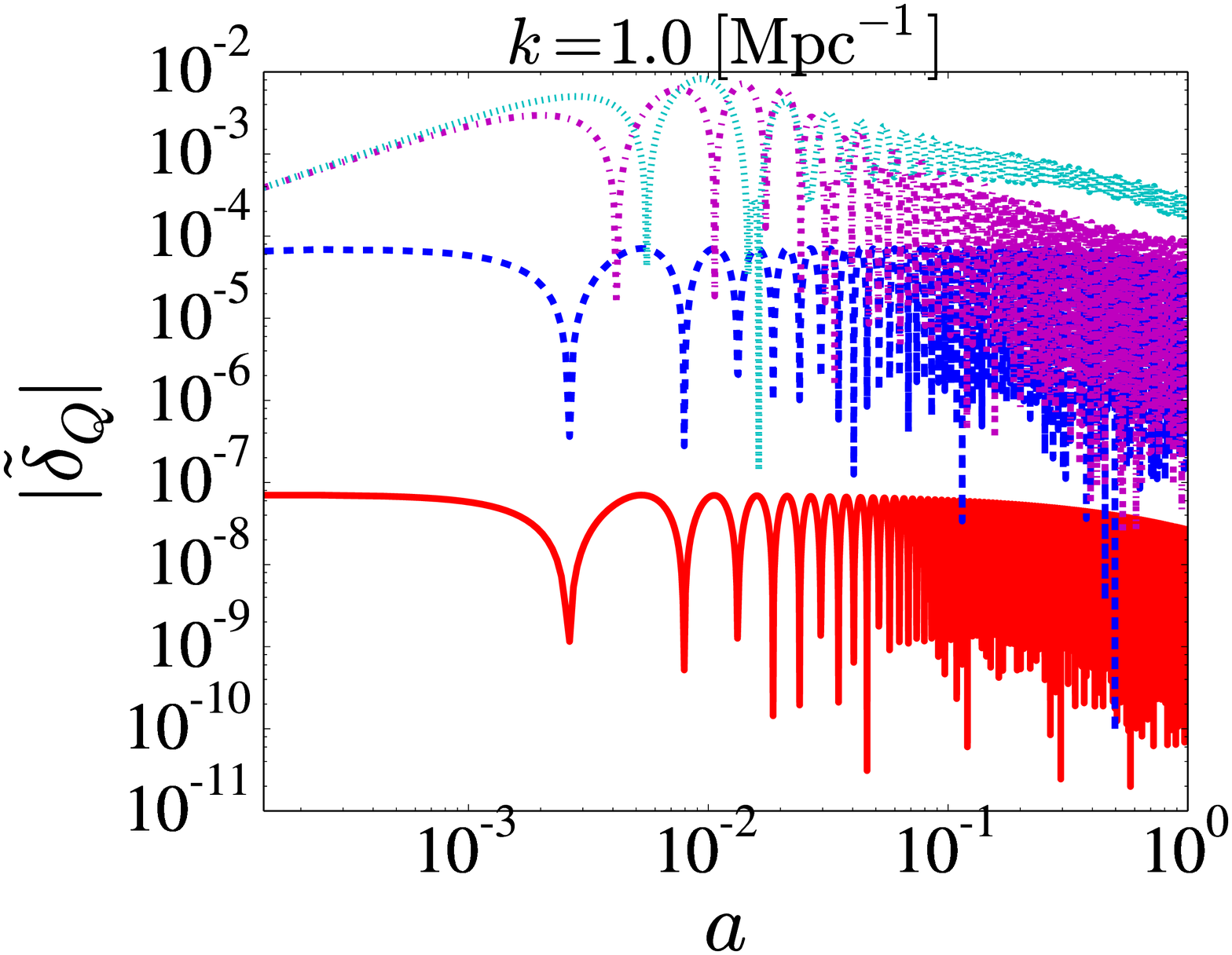} 
   \end{tabular}
  \caption{
    Evolutions of density perturbation of dark energy in Fourier space.  
   The evolution for $k=0.01, 0.1$ and $1.0$ Mpc$^{-1}$ 
    are shown in the left, middle and right panels, respectively.
    Each density contrast begins to oscillate when the Jeans length for each value of the speed of sound 
    crosses the wavelength.
    For calculation, we set $w = -0.9$ and $M=10^{15}M_{\sun}$.
      \label{fig:delta_q_k}}
\end{figure*}

In Fig. \ref{fig:delta_q_k}, we show 
the evolution of dark energy perturbation in Fourier space 
for $c_s^2 = 10^{-6}$, $10^{-4}$ and $10^{-1}$ and $w=-0.9$ 
for the $M=10^{15} M_{\sun}$ void.
We sample the fluctuations at $k=0.01, 0.1$ and $1$ ${\rm Mpc}^{-1}$, which
  correspond to larger, comparable and smaller scales, respectively, as compared to
  the void radius.
For $c_s^2=10^{-1}$, 
    $k>k_J$ for all three scales at $a<10^{-3}$.
   Therefore, the perturbation does not grow with time and
    gradually decays with oscillations.
 
  For $c_s^2=10^{-4}$, the fluctuation scale of $k=1.0$ ${\rm Mpc}^{-1}$ mode is always smaller than the Jeans
  length and thus shows the same trend as that for $c_s^2=10^{-1}$,
  whereas the scale $k = 0.1$ Mpc$^{-1}$ becomes closer to the Jeans length at $a = 10^{-2}$, 
  and thus, $\tilde{\delta}_Q(k=0.1 {\rm Mpc}^{-1})$ evolves
  with time only at an early epoch.  Once the scale crosses the Jeans length, it
  again shows a slight depression with oscillations. 
  
  For $c_s^2=10^{-6}$, $k<k_J$ for all three modes at $a<10^{-3}$,
  and thus $\tilde{\delta}_Q$ evolves with time at an early stage and it also
  decays with oscillation once the scale is close to the Jeans
  scale.  
  We also note that the
  amplitude of $\tilde{\delta}_Q$ is similar regardless of the value of the
  speed of sound when $\tilde{\delta}_Q$ is in the evolutional phase 
  (see $k=0.01$ Mpc$^{-1}$ at $a\ll 1$), whereas it is strongly dependent on
  $c_s^2$ once the scale of fluctuation reaches the Jeans length and
  $\tilde{\delta}_Q$ is constant or in an oscillationally decaying phase.

%%%%%%%%%%%%%%%%%%%%%%%%%%%%%%%%%%%%%%%%%%%%%%%%%%%%%%%%%%%%%%%%%%%%%
\section{Size distribution}
\label{sec:voidsize}
%%%%%%%%%%%%%%%%%%%%%%%%%%%%%%%%%%%%%%%%%%%%%%%%%%%%%%%%%%%%%%%%%%%%%
Dark energy perturbation also affects the size
distribution of voids
through the threshold of void
formation and the linear matter power spectrum. 
In this section, we revisit the size distribution of voids derived from
the excursion set theory, which was first applied to void formation by
\cite{Sheth:2004}, and then 
we extend the theory to the dark energy perturbation model.

%----------------------------------------------------------------------------
\subsection{Excursion set theory for voids}
\label{ssec:exvoid}
%----------------------------------------------------------------------------
Based on the Press-Schechter formalism \citep{Press:1974}, we consider a
 probability distribution for $\delta_M$,  which is the smoothed
linear density contrast at mass scale $M$ in Lagrangian space. 
Such a density contrast is expressed as 
\begin{align}
	\delta_M(\bm{q}) = \frac{3}{4 \pi R_L^3}\int _{|\bm{q}'-\bm{q}| \leq R_L} \delta(\bm{q}') d^3q',
\end{align}
where $\bm{q}$ is a Lagrangian coordinate, and $R_L$ is a smoothing scale in Lagrangian space 
corresponding to the 
mass scale via $M=4 \pi \bar{\rho}_m R_L^3/3=4 \pi \bar{\rho}_m(1+\delta_m) R_{\rm v}^3/3$
where $R_{\rm v}$ denote the size of voids in the Eulerian space.
If the density contrast exceeds a certain threshold $\delta_c$, 
the matter inside $R_L$ contributes to the formation of a halo of mass
more than $M$.   
The probability distribution that $\delta_{M}(\bm{q})$ exceeds
$\delta_c$ with mass between $M$ and $M+dM$
 can be interpreted as the fraction of mass between $M$ and $M+dM$ which contributes to the formation of a halo.  
 
In the excursion set theory (EST), 
we try to find the probability distribution of $\delta_M$ given 
the smoothing scale corresponding to mass $M$, where $\delta_M$
first crosses the density threshold \citep{Bond:1991,Zentner:2007}.
We assume that the probability distribution of the smoothed density fluctuation follows 
the Gaussian distribution with mean zero, and variance
\begin{align}
	S = \sigma^2(M) = \int \frac{k^2dk}{2\pi^2} \tilde{W}(kR_L) P(k).
	\label{eq:variance}
\end{align}
  
We define the probability distribution of $\delta_M$ on a scale $S$ for the random walk as
$\Pi(\delta_M,S)$.
Then, if we begin the trajectory from the horizon scale, we can set the initial condition   
$\Pi(0,0)=1$.  
We also assume that the transition probability is Gaussian: 
thus, the transition from  $ \Pi(\delta_M-\Delta \delta_M,S)$ to $\Pi(\delta_M,S+\Delta S)$ is  

\begin{align}
	\Pi(\delta_M,S+\Delta S)=  
	\int d(\Delta \delta_M) \frac{1}{\sqrt{2\pi \Delta S}} \exp \left(-\frac{(\Delta \delta_M)^2}{2 \Delta S} \right) \nonumber\\
	\times \Pi(\delta_M-\Delta \delta_M,S).
	\label{eq:transition}
\end{align}
Expanding the LHS of equation (\ref{eq:transition}) in terms of $S$ and the RHS in terms of $\delta_M$, 
we obtain the diffusion equation, 
\begin{align}
	\frac{\partial \Pi}{\partial S} = \frac{1}{2} \frac{\partial^2 \Pi}{\partial \delta_M^2}.
	\label{eq:diffusion}
\end{align}

As mentioned in \cite{Sheth:2004}, we set two barriers for void
  formation as a boundary condition for the random walk: linear
  density thresholds for halo and void formation.
For halo formation, $\delta_c=1.686$ is often used as the linear
  density threshold for collapsed objects in the literature, 
  while for void formation, $\delta_v=-2.717$ is
  analytically derived by using spherical model in the EdS universe \citep{Blumenthal:1992,Jennings:2013}.
This value is defined as the linear density contrast of a void shell crossing 
when the inner shell catches up with the outer shell: 
therefore the Lagrangian density at the shell diverges \citep{Sheth:2004}.
At that moment, the non-linear density contrast inside the void is
$\delta^{\rm TH}_m \simeq -0.8$. 

In order to obtain the void size distribution,
we need the probability of the first crossing of the void threshold
$\delta_v$ 
on a certain smoothing scale $R_L$ without  
crossing $\delta_c$ on any scales larger than $R_L$.
Any trajectories crossing $\delta_c$ before crossing
  $\delta_v$ correspond to the voids which are surrounded by high density
  regions, and then, eventually disappear during the
  structure formation (\textit{void in cloud}).
  Thus, we exclude those trajectories when we
  count voids.
  We consider only the first crossing of $\delta_v$ and do not count
  the voids any longer even if the trajectory crosses $\delta_v$ more
  than once at smaller scales in order to exclude the voids embedded in
  the larger void (\textit{void in void}) and to avoid the double counting.
  The solution of $\Pi(\delta_M,S)$ with above condition gives
  the probability distribution of $\delta_M$ between $\delta_v$ and $\delta_c$
  on a scale $S$ without crossing $\delta_c$ and $\delta_v$ on any $S'<S$.
  Thus the fraction of trajectories that $\delta_M$ satisfies the condition is 
  \begin{align}
  	F(S) = \int ^{\delta_c}_{\delta_v} \Pi(\delta_M,S) d \delta_M.
 \end{align}
 Let $f(S) dS$ be the probability that $\delta_M$ crosses either of the density 
thresholds between $S$ and $S+dS$, and such a probability is obtained by subtracting 
$F(S+dS)$ from $F(S)$.
Then, 
\begin{align}
	f(S)dS = - \frac{d F(S)}{dS} dS = \left. -\frac{1}{2} \frac{\partial \Pi}{\partial \delta_M}  \right|^{\delta_c}_{\delta_v} dS.
	\label{eq:first_crossing}
\end{align}
The limit of $\delta_M = \delta_c$ in the RHS of \rref{eq:first_crossing} is the probability that $\delta _M$ crosses $\delta_c$ 
for the first time between $S$ and $S+dS$,
whereas the limit of $\delta_M=\delta_v$ is the probability that $\delta_M$ reaches the value $\delta_v$.
Therefore, the probability that $\delta_M$ crosses $\delta_v$
for the first time between $S$ and $S+dS$ is
\begin{align}
	f(S)dS &=\left. \frac{1}{2} \frac{\partial \Pi}{\partial \delta_M}  \right|_{\delta_M=\delta_v} \nonumber\\
	 &=\sum_{n=1}^{\infty} \frac{n \pi D^2}{\delta_v^2} \sin (n \pi D) \exp \left( - \frac{n^2 \pi^2 D^2}{S} \frac{S}{\delta_v^2} \right) dS,
	\label{eq:first_crossing_deltav}
\end{align}
where $D = |\delta_v|/(\delta_c + |\delta_v|)$. 

The mass function 
is expressed as 
\begin{align}
	\frac{d N}{d \ln M } = \frac{\bar{\rho}}{M} S f(S)\frac{d \ln S}{ d \ln M},
\end{align}
where $N$ is the number density of voids including mass $M$.
It is useful to characterise the void by its size instead of
  mass. Thus, we rewrite the abundance
  of voids in terms of the size by using the relation,
\begin{equation}
	\frac{d N}{ d \ln R_{\rm v}}=\frac{d N}{ d \ln R_L}=3 \frac{d N}{ d \ln M}.
	\label{}
\end{equation}
Finally, we obtain the void size distribution in Eulerian space,
\begin{equation}
	\frac{dN}{dR_{\rm v}} = (1+\delta_m)^{\frac{1}{3}} \left( \frac{3}{4 \pi R^3_L} \right) f(\nu) \frac{d \nu}{d R_L},
	\label{eq:abundance_size}
\end{equation}
where $\nu = \delta_v^2/S$.

 We adopt an approximate expression for $f(\nu)$ \citep{Sheth:2004},
\begin{equation}
	\nu f(\nu)\approx \sqrt{\frac{\nu}{2\pi}} \exp \left(-\frac{\nu}{2} \right) 
	\exp \left( -\frac{| \delta_v |}{\delta_c} \frac{ D^2}{4 \nu}-2\frac{D^4}{\nu^2} \right).
	\label{eq:f_nu}
\end{equation}

The dark energy perturbation affects the number of void in two ways.  
First, the dark energy perturbation affects $\nu$ through the
  variance, equation (\ref{eq:variance}).  
  We calculate the linear power spectrum $P(k)$ using
  the publicly available code, \texttt{CAMB} \citep{CAMB} to derive
  $S$ for the dark energy perturbation model.
  Second, the dark energy  perturbation affects the density threshold for
  cosmic voids.
  We solve  equations (\ref{eq:continuous_deltam_lin_k}),
  (\ref{eq:Euler_deltam_lin_k}) and (\ref{eq:2nd_deltade_k})
  to find the linear density threshold corresponding to the non-linear
  density fluctuations of $\delta_m = -0.8$ \citep{Blumenthal:1992}.

\subsubsection{Dependence on $c_s^2$}
\begin{figure*}
  \begin{tabular}{cc}
    \includegraphics[width=0.5\linewidth]{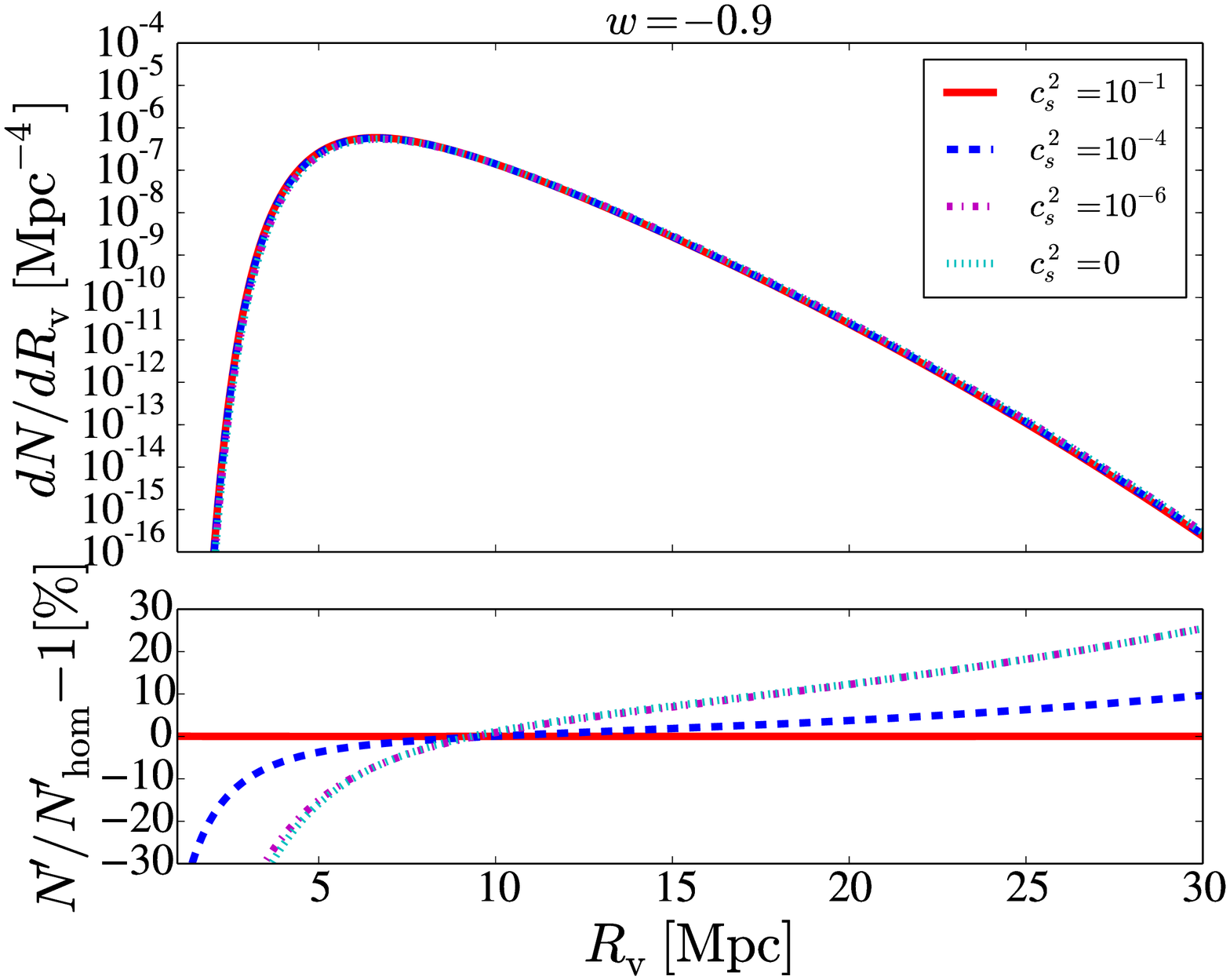} &
    \includegraphics[width=0.5\linewidth]{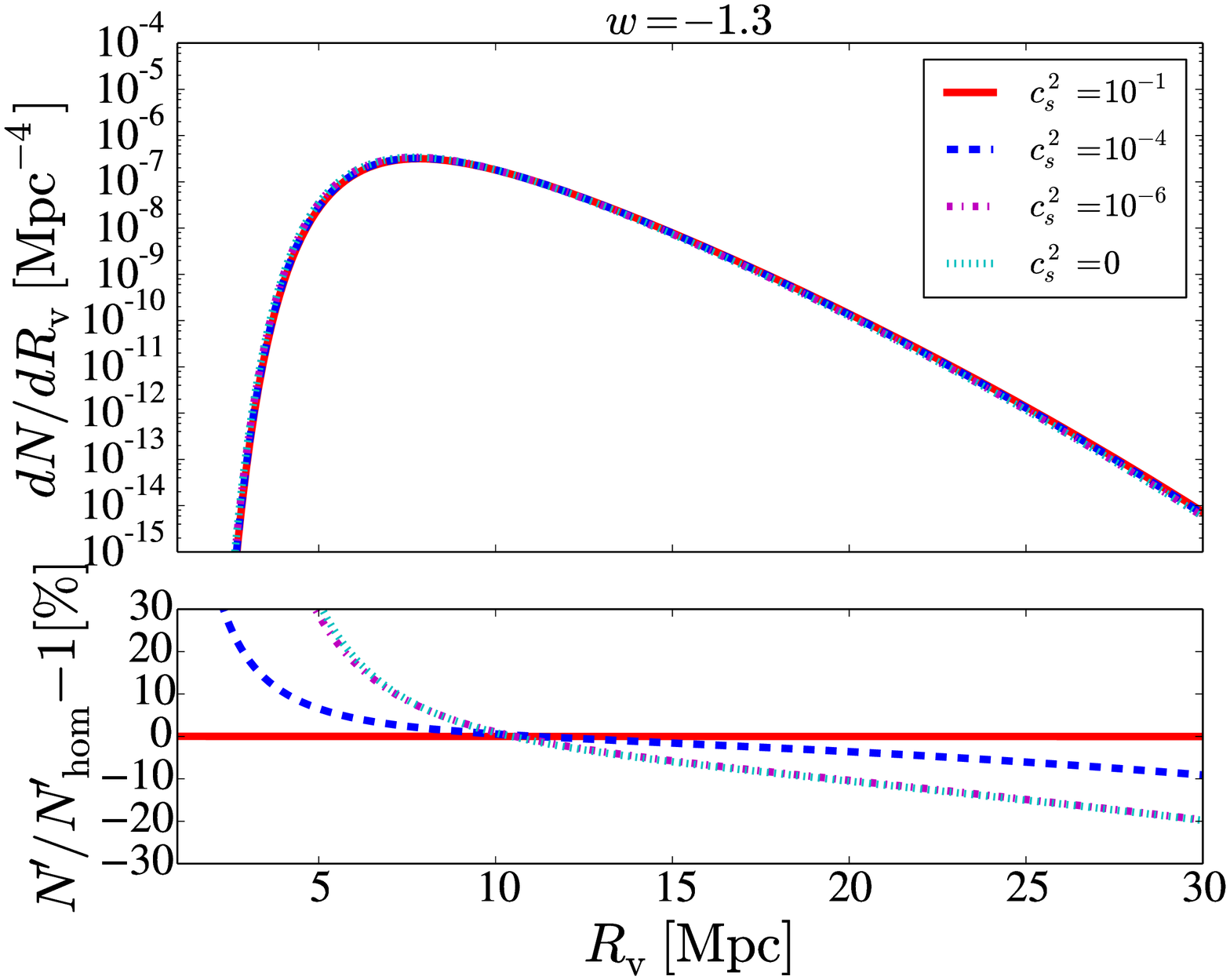}
  \end{tabular}
  \caption{Comparison of the abundance of voids at present time.
      We fix $w = -0.9$ (left) and $w=-1.3$ (right), and the values
      of speed of sound as $c_s^2 =0, 10^{-6}, 10^{-4}, 10^{-1}$.
      The lower panels show the fractional difference from the homogeneous
      dark energy model with the same equation of state parameter,
      where $N' = dN/dR_{\rm v}$.
      For both $w$, the deviation from  the case for homogeneous dark
      energy is noticeable at larger radii: more than $10 \%$ for
      $c_s^2=0$ for 30 Mpc voids.
     \label{fig:abundance_cs}}
\end{figure*} 

\begin{figure}
	\includegraphics[width=1\linewidth]{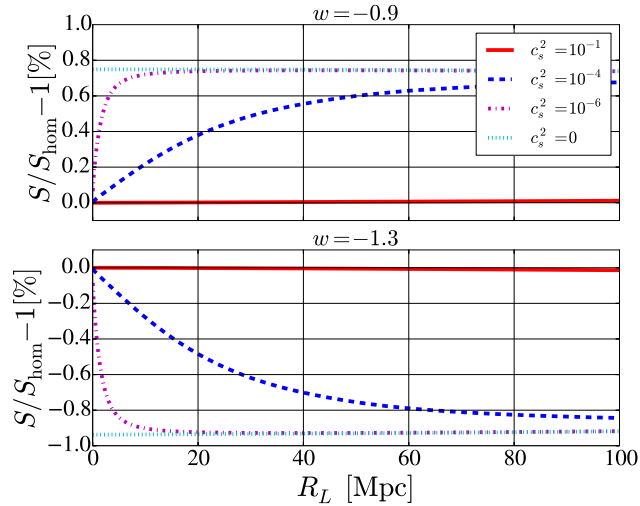}
	\caption{Comparison of variance at present time.   We set $w=-0.9$ for 
	the upper panel and $w=-1.3$ for the lower panel.  Both panels show the 
	fractional difference from the homogeneous dark energy model.  }
	\label{fig:variance}
\end{figure}

 \begin{figure*}
  \begin{tabular}{cc}
    \includegraphics[width=0.5\linewidth]{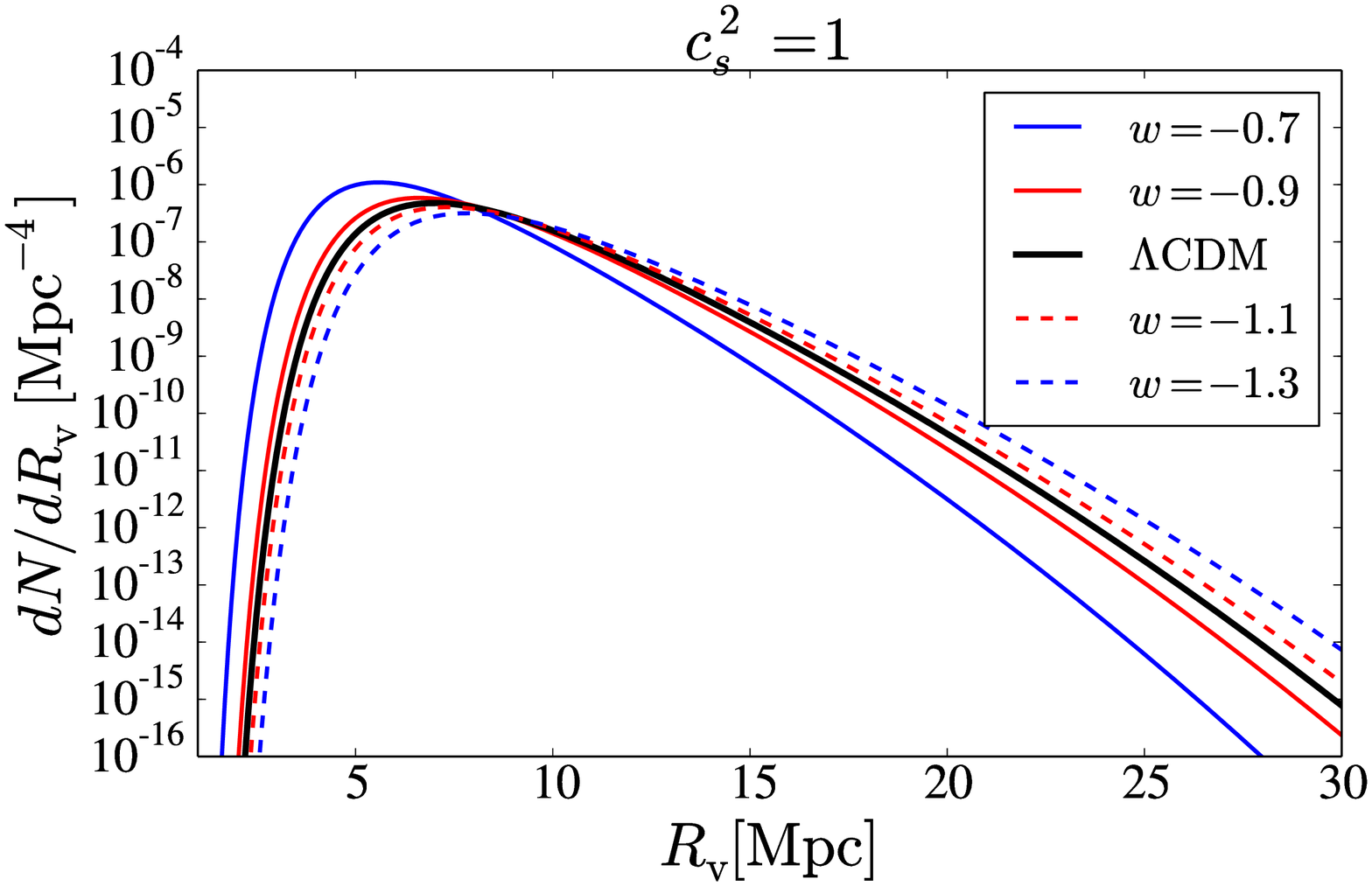} 
    \includegraphics[width=0.5\linewidth]{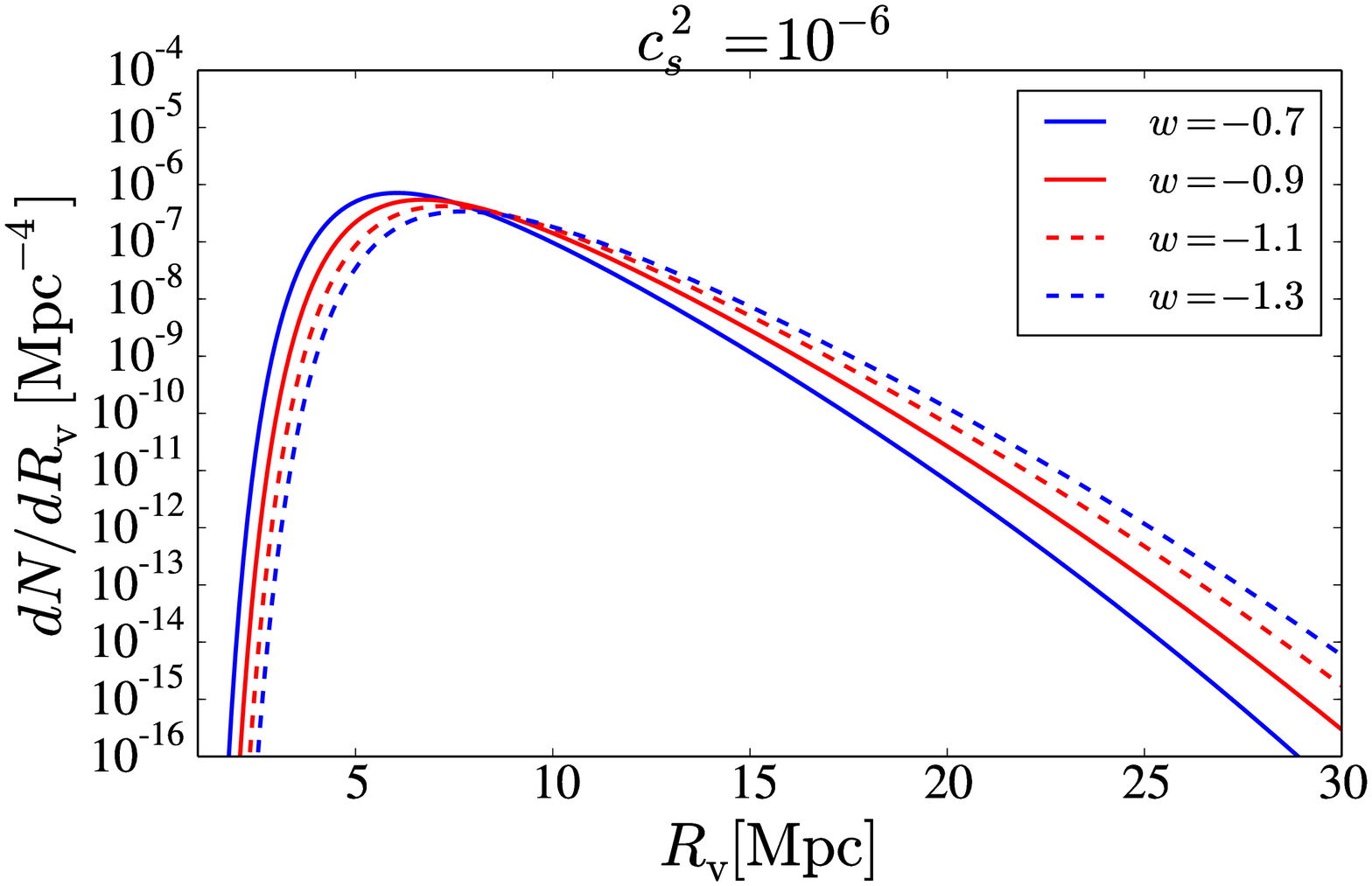}
  \end{tabular}
  \caption{ 
     Comparison of the abundance of voids for different equation of
      state parameters for fixed speed of sound. We set $c_s^2=1$ (left)
      and $c_s^2=10^{-6}$ (right). Models with $w<-1$ enhance the
      number of voids with large radii whereas they suppresses the number of
      smaller voids. These effects are more prominent in the absence
      of dark energy perturbation.
    \label{fig:abundance_w}}
\end{figure*}

%----------------------------------------------------------------------------
\subsection{Results}
\label{ssec:results_abundance}
%----------------------------------------------------------------------------
In this section, we demonstrate the numerical results for the
  effect of the dark energy perturbation on void abundance.

Figure \ref{fig:abundance_cs} shows the size distribution of voids at the
present time.
The size distribution does not show any differences between 
$c_s^2=10^{-1}$ and the homogeneous dark energy model ($c_s^2=1$) for both 
$w=-0.9$ and $-1.3$.
For $w=-0.9$ and $c_s^2<10^{-4}$,  the number of large voids with a radius of $30$ Mpc 
is larger by more than 10 \% compared with the homogeneous dark energy
model, whereas small voids with radii smaller than $10$ Mpc decrease.

For $w=-1.3$, the dark energy perturbation works in a manner opposite to that for the
  $w=-0.9$ model; the dark energy perturbation suppresses the number of
  large voids and enhances the number of small voids.
  For $c_s^2<10^{-4}$, the number of voids of radius of $30$ Mpc is
  suppressed by also more than $10\%$ as compared with that in the homogeneous dark energy
  model.  

We  see almost no difference in void abundance 
between $c_s^2 = 0$ and $10^{-6}$.
As described in Section \ref{sec:results_size}, the Jeans length for $c_s^2=10^{-6}$ is
sufficiently smaller than the void radius, and therefore, large
voids are affected by the dark energy perturbation in a similar manner
in the $c_s^2=0$ and $10^{-6}$ models.

    According to the results of Section \ref{sec:results_size}, 
  for individual voids with radius $R_{\rm v}\sim 30 \ {\rm Mpc}$ at present time ($a=1$), 
  the deviation of the radius from the homogeneous dark energy model is of the order of 0.1\%.  
  In contrast, the deviation of the abundance from that for the homogeneous dark energy model is of the oder of 10\%. 
  The difference in void abundance mainly arises from the difference in variance $S$.  
  As shown in Fig. \ref{fig:variance}, for $w=-0.9$, the deviation in $S$ with $c_s^2=0$ 
  from the homogeneous dark energy model at a smoothing scale $R_L$ = 30 Mpc is about 0.75 \%,
  while for $w = -1.3$ such deviation is -0.94\%. 
  In addition, the size function includes the exponential, and therefore, the deviation 
  in the size function is significant.  
  
  We also see drastic deviations at smaller scales which have the opposite    
  trend for larger scales. For larger scales, the first exponential term in \rref{eq:f_nu} mainly contributes to 
  the abundance of voids.  
  For $w=-0.9$ we see that the variance with dark energy perturbation 
  is enhanced as compared with the homogeneous dark energy model,  
  indicating that the argument of the exponential approaches $0$ when there is dark energy perturbation
  which leads to an increase in the abundance of large voids.
  In contrast for small scales, the second exponential term contributes to the 
  abundance and the argument of the exponential rapidly drops
  in the presence of dark energy perturbation, suppressing the abundance of small voids.
  
  For $w=-1.3$ the difference in variance between the inhomogeneous and homogeneous dark energy models 
  have opposite signs as compared with the case of $w=-0.9$.
  Thus, the abundance of voids is enhanced on smaller scales 
  and suppressed on larger scales in the presence of dark energy perturbation.  
    
    We also confirm 
that the peak of the abundance is approximately at $R = 5\sim10$ Mpc.   
Furthermore, we find that different values of the speed of sound lead to different 
peak locations in the size function.
Compared with the $c_s^2=0$ case, 
the position of the peak shifts to larger scale by 1.3\%
whereas it shifts smaller scale by 1.5\% for $w = -1.3$ when $c_s^2 = 10^{-1}$.

\subsubsection{Dependence on $w$}

Figure \ref{fig:abundance_w} shows the void abundance for a fixed
  value of the speed of sound but different equation of state.
  We see that $w<-1$ models increase the number of voids with radius larger than $10$ Mpc, 
   while  $w>-1$ models  increase that of voids on scales smaller than $10$ Mpc. 
  These enhancements are smaller for the inhomogeneous dark energy model 
  than for the homogeneous dark energy model.
For example, the number of voids with $R_{\rm v}=30$ Mpc for $w=-1.3$ is 
  about 1500 times larger than that for $w=-0.7$ for the homogeneous dark
  energy model, whereas  for the $c_s^2=10^{-6}$ model it is only about 280 times larger.
  
  As we have seen in previous sections, the sign of dark energy perturbation 
	is different depending on whether $w$ is greater or less than $-1$.  
	Therefore, when $w$ is less negative, the dark energy perturbation 
	causes an increase in the number of large scale voids while it leads to a decrease in that of the small scale voids.
	In contrast, when $w$ is more negative, dark energy perturbation affects in an 
	opposite manner to the case of $w>-1$ so that the number of large voids is suppressed 
	while that of small voids is enhanced.

%%%%%%%%%%%%%%%%%%%%%%%%%%%%%%%%%%%%%%%%%%%%%%%%%%%%%%%%%%%%%%%%%%%%%
\section{Summary}
\label{sec:summary}
%%%%%%%%%%%%%%%%%%%%%%%%%%%%%%%%%%%%%%%%%%%%%%%%%%%%%%%%%%%%%%%%%%%%%  
We  investigated 
the effects of dark energy perturbation on the formation 
of cosmic voids. 
We treated the speed of sound and the equation of state of dark energy 
as constant parameters in our model.
We studied
the dependence of the formation of an isolated spherically symmetric void
on these parameters and the initial size of the void.  
We found
that the effects of the different values of the speed of sound and initial sizes are much small. 
These results are broadly consistent with those of \cite{Novosyadlyj:2016},
and may lead us to the conclusion
that the dark energy perturbation does not greatly affect void formation.

We also 
investigated the effects of the dark energy perturbation on
the abundance of voids based on the EST. 
We found that the differences between the homogeneous  
and inhomogeneous dark energy models are significant when the speed of sound is
much smaller than that of light. 

For both $w =-0.9$ and $w=-1.3$, 
the difference in the size of voids with an initial mass of $10^{15} M_{\sun}$ 
between the cases of $c_s^2=1$ and $c_s^2={0}$ is of the order of 0.1\% at present time.
In contrast, the difference in the abundance of the corresponding voids is more than 20 \%.
As the size function has an exponential tail at larger scales, 
a subtle change in the void size may cause a drastic amplification
of the void abundance.

%%%%%%%%%%%%%%%%%%%%%%%%%%%%%%%%%%%%%%%%%%%%%%%%%%%%%%%%%%%%%%%%%%%%%
\section*{Acknowledgements}
%%%%%%%%%%%%%%%%%%%%%%%%%%%%%%%%%%%%%%%%%%%%%%%%%%%%%%%%%%%%%%%%%%%%%
We acknowledge the use of publicly available code \texttt{CAMB}. 
This work is in part supported by MEXT KAKENHI Grant Number 16H01096, 16H01543
and MEXT's Program for Leading Graduate Schools PhD professional,``Gateway to Success in Frontier Asia".  
AN is gratefully acknowledge the IAR grant at Nagoya University.

\bibliographystyle{mn2e}

\end{document}